# Electromechanically Programmable Space-Time-Coding Digital Acoustic Metasurfaces

*Hamid Rajabalipanah[+], Mohammad Hosein Fakheri[+], Ali Abdolali[*]*

Applied Electromagnetic Laboratory, School of Electrical Engineering, Iran University of Science and Technology, Tehran, 1684613114, Iran

[+]These authors have contributed equally to this work.
E-mail: abdolai@iust.ac.ir

## Abstract

Over the last couple of years, the digital coding acoustic metasurfaces have been developed rapidly as a highly active research area for their unique and flexible manipulation of acoustic wavefronts. Nevertheless, all recent attentions in the acoustic community have been mainly concentrated on space-encoded architectures, leaving the room free for benefiting from the unique features of spatiotemporally modulated metasurfaces. By entering the world of time, here, we propose a space-time-coding acoustic digital metasurface with exotic ability to dynamically transfer the energy of the carrier acoustic signal to a series of harmonic components, with equivalent magnitudes and phases that can be precisely and independently engineered. The contributing elements are composed of a straight pipe, and four shunted Helmholtz cavities (HCs) with transmission phases dynamically controlled by exploiting a high-speed electromechanical actuation system. Several illustrative examples have been presented to demonstrate that by distributing the coding sequences in both space and time dimensions, diverse scattering functionalities can be elaborately acquired for one or multiple harmonic frequencies in a programmable way. Numerical and theoretical results are in an excellent agreement, thereby elucidating that this next generation of programmable acoustic metasurfaces, without restoring to high-cost nonlinear components, opens up unprecedented potential for efficient harmonic control used in adaptive beamforming and acoustic imaging systems.

## 1. Introduction

Recent years have witnessed a steadily increasing interest in exploring the use of metasurfaces, two-dimensional (2D) equivalence of metamaterials, for unprecedented acoustic wave manipulations, toward potential applications in a wide variety of scientific and engineering disciplines such as medical ultrasound and sound communication[1,2,3,4,5,6,7,8,9,10]. With much better integrability and lower insertion losses, acoustic metasurfaces have demonstrated exotic capabilities to manipulate sound properties by locally controlling the amplitude[11] and phase[12,13,14], of subwavelength meta-atoms in a predefined and desired manner. Impressing abrupt phase discontinuities to the incident acoustic fields has given rise to a wealth of unique functionalities, including acoustic cloaking[15], acoustic holography[3], and wave-based analog computing[16].

Among multitudinous researches in the land of acoustic wave manipulation, coding metasurfaces pioneered by Cui *et al.*[17], have become an essential direction through constructing "metasurface bytes" with proper spatial mixtures of "digital metasurface bits". Combination of the digital world and physical metamaterials has been revolutionary suggested to be an alternative wave manipulation route for the traditional design tools such as acoustic[18,19,20,21] and electromagnetic [22,23,24] coordinate transformations. Shortly after the advent of digital metasurfaces in the microwave[25,26,27,28] region, the idea has permeated into acoustics[29] by Xie *et al.* to greatly facilitate the design and optimization processes and create more diverse functionalities. The essence of the acoustic coding metasurfaces is to macroscopically arrange only finite (typically two) kinds of sub-wavelength meta-atoms with "0" and "1" digitals states standing for the phase responses of 0 and π, respectively[25]. Conventionally, by exploiting the space-coiling structures[30,31] and Helmholtz resonators[32,33], the majority of designs have dealt with acoustic coding elements of fixed geometries optimized for a particular appeal[29,34,35,36,37,38]. Indeed, the metasurface mission remains unchanged after being fabricated.

Up until now, reconfigurable architectures such as origami-inspired metamaterials[28] and elastomeric helices[29] have been demonstrated to tailor the acoustic wavefronts[39]. More recently, Tian *et al.* reported a water-based programmable acoustic metasurface that contains an array of tunable subwavelength unit cells to realize multiple 2D wave manipulation functions[40]. Although programmable acoustic metasurfaces are experiencing a strong surge of interest, the current proposals are restricted to space-coding scenarios[41,42,43] in which the time evolution has not been given into consideration. Particularly, the coding sequence is generally fixed in time, and is changed by the control system only to switch the functionalities whenever needed. The systems with time-varying physical parameters bring many interesting applications that cannot be achieved by the traditional programmable metasurfaces such as breaking time-reversal symmetry and Lorentz reciprocity[44,45] and harmonics generation[46]. Despite all recent efforts to foster the space-time-coding digital metasurfaces in the electromagnetic community[47,48,49,50,51,52,53], the successful demonstration in acoustics has been scarce, leaving the room free for benefiting from the unique features of spatiotemporally modulated metasurfaces in control of multiple sound harmonics including the beam directions and intensities. The few existing studies in this research area were also based on analog modulations, which are sophisticated to implement in practice[54,55,56,57,58].

In this paper, we propose the first design of time-domain digital coding metasurfaces to extend the arsenal of metasurface-based acoustic wave manipulations. The designed transmission-type coding elements are electromechanically actuated shunted Helmholtz resonators whose acoustic operational states can be dynamically switched among "0" and "1" digital modes with an engineered modulation speed. Specifically, a set of coding sequences are switched cyclically in a predesigned time period to produce and control nonlinear responses in free space without any need to nonlinear materials. Firstly, we outline the fundamental theory of time-varying acoustic coding metasurfaces and then present several illustrative examples to demonstrate the potential of our approach in accurate control of both amplitude and phase distributions for all emerging harmonics. Meanwhile, extra time delays are imposed on the modulation signal of the meta-atoms to create a metasurface, enabling wavefront reshaping for high-order harmonics. The proposed approach expands the scope of the applications for the acoustic metasurfaces significantly, and holds great potential for future imaging and beamforming scenarios.

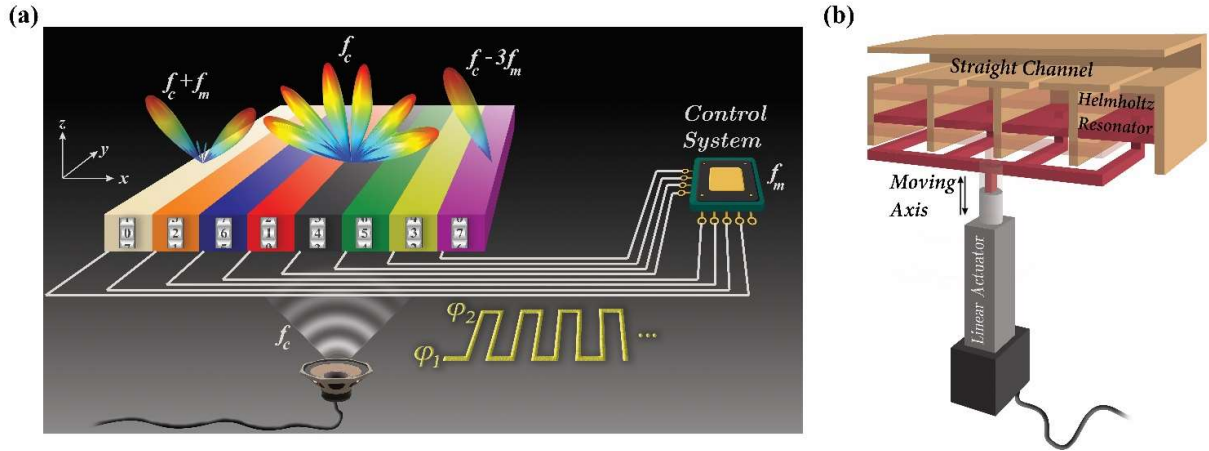

**Figure 1.** (a) Conceptual illustration of the nonlinear acoustic wave manipulation for real-time scattering pattern control in both space and frequency domains based on a space-time-coding digital metasurface comprising of (b) electromechanically programmable acoustic meta-atoms. Once the cavity size of the acoustic meta-atom is altered by means of the linear actuator, the transmission phase can be dynamically switched between different states while maintaining high energy transmission.

## 2. Space-time-coding Strategy

### 2.1. Electromechanically Programmable Coding Elements

The designed building unit and the overall configuration of the proposed transmission-type space-time modulated acoustic metasurface are depicted in **Figures 1a, b**. The programmable coding meta-atoms consist of a straight pipe, a series of four subwavelength shunted Helmholtz cavities (HCs) embedded in air, and subwavelength slits (see **Figure 1b** and **Figure 2a**). Acting as lumped elements, the series connection of cavities aids to modulate the wave number $k_{eff}$ or equivalently, the transmission phase of the coding meta-atom while the combination of cavities, slits, and pipes provides hybrid resonances that overcome the impedance mismatch between the resonators and the surrounding air for high transmission. The working (carrier) frequency $f_c$ of the acoustic metasurface is set near 3 kHz. The thickness of the acoustic meta-atom is optimized to have a propagation distance $w$=50 mm (0.45 $\lambda_0$ corresponding to $f_0$= 3 kHz). The fixed geometrical parameters are h=18.3 mm, $h_1$=5 mm, $d_1$=10.8 mm, $d_2$=1 mm, and p=12.2 mm. Through changing the cavity height, $h_x$, the coding element has the ability to span the whole transmission phase range of 0 to 2π. To grasp a complete picture of the transmission spectra for the optimized meta-atom, the numerical simulations were performed on COMSOL Multiphysics 5.4, a commercial finite element package with a pressure acoustics module. Sound hard boundaries have been applied to the lateral walls, and the background medium is air with the mass density $\rho_0$=1.21 kg/m$^3$ and speed of sound $c_0$=343 m/s. The material used for constructing the acoustic elements is a high stiffness UV curable acrylate polymer with much larger impedance than that of air, allowing us to assume that the walls of the resonators are acoustically hard. The phase and amplitude responses of acoustic waves passing through the designed meta-atom are plotted in **Figures 2b, c** for different cavity heights. As can be seen, when the cavity height increases from 4 to 10 mm, an almost full 2π span of transmission phase change can be achieved in the vicinity of 3.0 kHz. Meanwhile, when $h_x$ lies between 6.5 mm and 9.5 mm, the transmission amplitude remains above 87%, ensuring a high energy throughput. From the results, one can immediately deduce that the "$\varphi_2$=0" and "$\varphi_1$=π" relative transmission phases can be dynamically addressable through switching the cavity size of each coding element between $h_{x1}$=6.7 mm and $h_{x2}$=9 mm. For the ease of description, the coding elements with the "0" and "π" transmission phases are denoted by "0" and "1" digital bits. The space-coding and time-coding sequences stipulate the manner of "0" and "1" combination in the temporal and spatial domains, respectively. **Figure 1a** illustrates our idea to design a programmable acoustic meta-atom in which, *electromechanical actuation*, is chosen as the key mechanism for real-time cavity height modulation. The size of each HC can be independently altered by using a high-speed electrically controlled mechanical linear actuator that typically operates by conversion of rotary motion into linear motion along a straight line[59]. The linear actuators are mechanically connected to the bottom part of the contributing HCs, providing the required displacement for making a pre-determined change in the cavity height, and so, a phase jump in the transmitted acoustic fields. The setup can be implemented by a motor assembled with the necessary mechanical components and the peripheral programming circuit. A high-speed microcontroller can elaborately serve to produce the desired alternating current (AC) control signals with different voltages, periods and time delays based on the time-varying coding sequences already saved in its memory. The switch time depends on several structural and environmental parameters of the linear actuators, but, one can expect a typical value of $t_r$=0.6~400 μs for "0→π" or "π→0" transmission phase toggling.

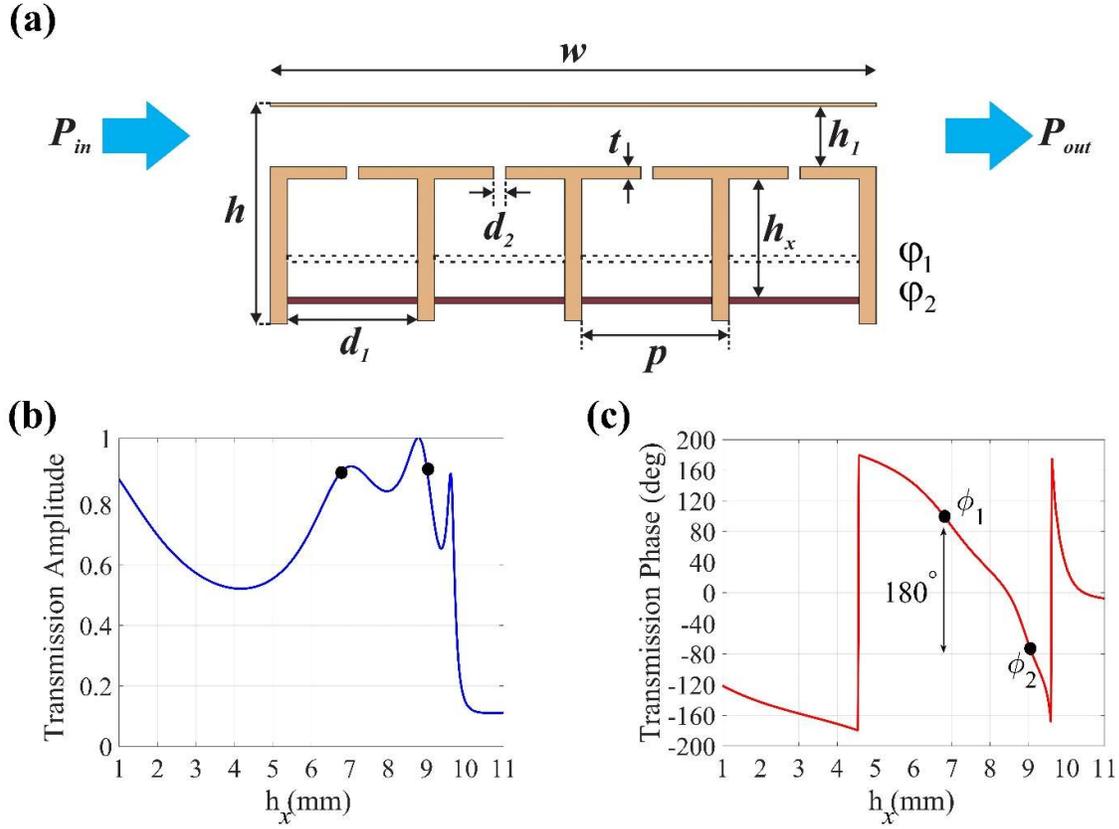

Figure 2. (a) The front view of the designed acoustic meta-atom, where the incident sound propagates along the positive z-direction. The influence of the cavity height variation on the (b) transmission phase and (c) amplitude responses of the electromechanically programmable meta-atom at $f_c$=3 kHz.

## 2.2. Fundamental Theory

**Figure 1a** sketches a space-time-coding acoustic metasurface with M×M electromechanically programmable elements. The transmission phase responses of all coding elements are periodically modulated in time between two different values of $\varphi_n=(\varphi_2-\varphi_1)q+\varphi_1$, q=0, 1. $T_m$ represents the period of time modulation. Hence, a time-coding sequence including "0" and "1" digits along with a pre-defined pair ($\varphi_1$, $\varphi_2$) can describe the phase level of the contributing elements at each time interval. In this section, to observe the nonlinear generation capability of the designed metasurface, the interaction between acoustic waves and space-time-coding metasurface is analyzed. Let us consider that the programmable metasurface of **Figure 1a** is excited by a monochromatic acoustic plane wave with the time-harmonic form $P_i(t)=P_0 exp(j\omega_c t)$. According to the time-switched array theory, the transmitted pressure field of the $(u, v)^{th}$ element can be expressed as

$$P_t(t) = \xi_{uv}(t) P_i(t) \tag{1}$$

Here, we assume that the modulation frequency, $f_m=1/T_m$, is far less than the carrier frequency, $f_c$, allowing us to adiabatically extend the approximate modelling already introduced for space-coding metasurfaces[60,61]. Moreover, $\xi_{uv}(t)$ is a periodic signal of length L denoting the complex transmission coefficient of the $(u, v)^{th}$ element in each time interval. Therefore, it can be written as the linear combination of $L$ scaled and shifted pulses $\xi_{uv}(t) = \sum_{n=1}^{L} \xi_{uv}^n \Pi^n(t)$, $0<t<T_m$, in which

$$\Pi^n(t) = \begin{cases} 1 & (n-1)T_m/L < t < nT_m/L \\ 0 & elsewhere \end{cases} \quad (2)$$

wherein, $\xi_{uv}^n \in \{e^{j\varphi_1}, e^{j\varphi_2}\}$ indicates the transmission coefficient of the $(u, v)^{th}$ element during the $n^{th}$ interval, i.e., $(n-1)T_m/L < t < nT_m/L$, with $\varphi_1$ and $\varphi_2$ representing the two available transmission phases. Using Fourier series expansion, we have[47]

$$\xi_{uv}(t) = \sum_{m=1}^{\infty} a_{uv}^{mn} \exp(jm\omega_m t) \quad (3)$$

$$a_{uv}^m = \frac{1}{T_m} \int_0^{T_m} \xi_{uv}(t) \exp(-jm\omega_m t) dt = \frac{1}{T_m} \sum_{n=1}^{L} \xi_{uv}^n \int_0^{T_m} \Pi^n(t) \exp(-jm\omega_m t) dt$$

$$= \frac{1}{T_m} \sum_{n=1}^{L} \xi_{uv}^n \int_{(n-1)T_m/L}^{nT_m/L} \Pi^n(t) \exp(-jm\omega_m t) dt = \sum_{n=1}^{L} \xi_{uv}^n \frac{\sin(m\pi/L)}{m\pi} \exp\left[-j\pi m(2n-1)/L\right] \quad (4)$$

The Fourier Transform of Eqs. (1), (3) yields

$$\xi_{uv}(\omega) = \sum_{m=-\infty}^{\infty} a_{uv}^{mn} \delta(\omega - m\omega_m) \quad (5)$$

$$P_t(\omega) = 2\pi \xi_{uv}(\omega) * P_i(\omega) = 2\pi \xi_{uv}(\omega - \omega_c) = 2\pi \sum_{m=-\infty}^{\infty} a_{uv}^{mn} \delta(\omega - \omega_c - m\omega_m) \quad (6)$$

As expected, modulating the transmission phases with AC signals yields a series of harmonics with intensities determined by the phase pair ($\varphi_1$, $\varphi_2$) in Eq. (4). The spectral position of the emerging harmonics is merely specified by the period of time modulation. Herein, $a_{uv}^m$ refers to the equivalent transmission coefficient of the $(u, v)^{th}$ element at the $m^{th}$ harmonic frequency ($f_c+mf_0$) which can be elaborately engineered via a suitable choice of the time-coding sequence. For better understanding the presented coding scheme, we consider a more straightforward scenario in which a time-domain square wave signal with $\exp(j\varphi_1)$ and $\exp(j\varphi_2)$ levels ($L=2$) is exploited to drive the programmable elements. Thus, Eq. (4) is reduced to

$$a_{uv}^{m} = \begin{cases} \cos\left(\dfrac{\phi_1 - \phi_2}{2}\right) \measuredangle \dfrac{\phi_1 + \phi_2}{2} & m = 0 \\ \dfrac{2}{m\pi}\sin\left(\dfrac{\phi_1 - \phi_2}{2}\right) \measuredangle \dfrac{\phi_1 + \phi_2}{2} & m = \pm 1, \pm 3, \pm 5,... \\ 0 & m = \pm 2, \pm 4, \pm 6,... \end{cases} \quad (7)$$

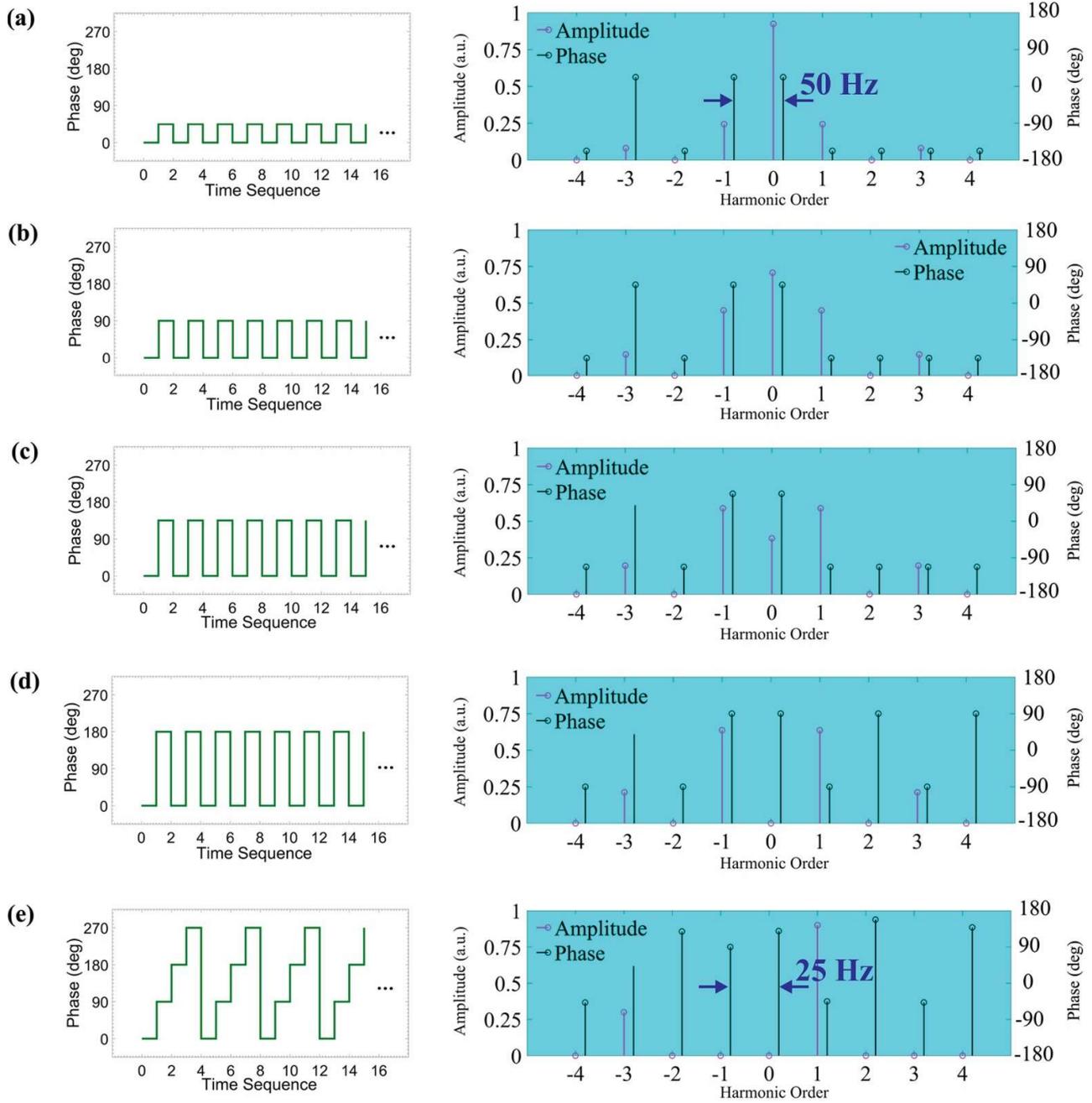

**Figure 3.** The harmonic phases/intensities distribution of the time-domain digital acoustic metasurface at 3 kHz with the pulse duration $\tau$=10 ms under different time-varying coding sequences with (a) $(\varphi_1, \varphi_2) = (0, 45°)$, (b) $(\varphi_1, \varphi_2) = (0, 90°)$, (c) $(\varphi_1, \varphi_2) = (0, 135°)$, (d) $(\varphi_1, \varphi_2) = (0, 180°)$, and (d) $(\varphi_1, \varphi_2, \varphi_3, \varphi_4) = (0, 90°, 180°, 270°)$.

The synchronous component (m = 0) and the odd harmonics are found effective in the transmitted acoustic waves due to the Fourier properties of the square wave signals. **Figures 3a-d** show the harmonic intensities/phases distribution of the time-modulated digital metasurface, where different pairs ($\varphi_1$, $\varphi_2$) and time-domain sequences are considered. According to the switching speed range of the electromechanical controlling system, the pulse width is selected as $\tau$=10 ms to result in the frequency gap of 50 Hz between different harmonics. As can be observed, the change in the phase states $\varphi_1$ and $\varphi_2$ dramatically affects the equivalent phase/intensity distribution at different harmonic frequencies. In accordance with Eqs. (4), (7), by approaching $\varphi_1-\varphi_2$ to the anti-phase condition, the zeroth harmonic is noticeably attenuated since most of the transmitted energy is shifted into the other harmonic frequencies. In this case, the two symmetric components m= ±1 occupy the 81% of the overall transmission energy. In addition, the amplitude spectra have a mirror-type intrinsic symmetry with respect to the carrier frequency, since the transmission coefficient of elements is real in the time domain. To break such symmetry, we can exploit from those time-coding sequences with more than two switch levels of anti-symmetric phase ramps to eradicate the unwanted harmonics and generate asymmetric power distributions among corresponding positive and negative harmonic frequencies. In this case, the asymmetric spectral response is the result of violating $a^{-m}=(a^m)^*$ owing to the fact that the transmission coefficient of the acoustic elements is no longer real in the time domain. As demonstrated in **Figure 3e**, a 2-bit time-coding sequence with $\varphi \in \{0, \pi/2, \pi, 3\pi/2\}$ yields an asymmetric spectral response with intensities directly controlled by the cyclicity of the contributing phase states. Owing to the pre-defined pulse width, the frequency space between the neighbouring components is 25 Hz.

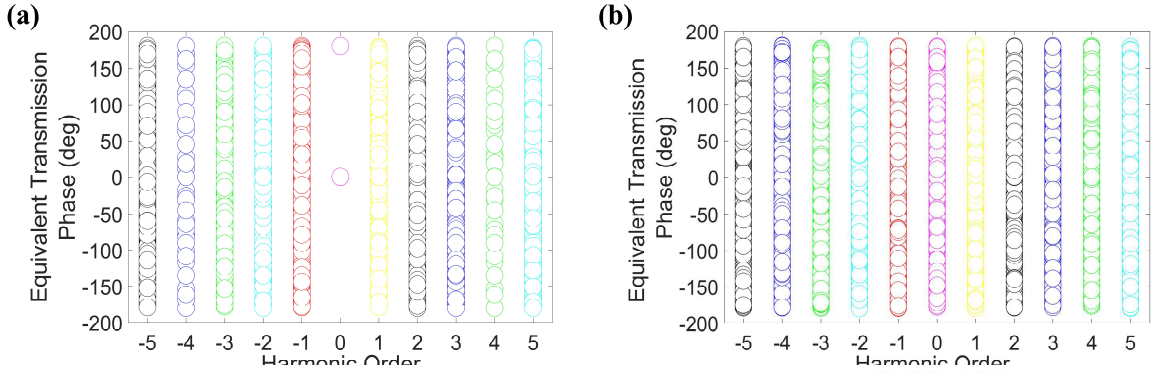

**Figure 4. Equivalent phase coverages of the time-domain digital acoustic metasurface at different harmonic frequencies. The transmission phase of the acoustic elements is periodically switched between (a) ($\varphi_1$, $\varphi_2$) = (0, 180°) and (b) arbitrary ($\varphi_1$, $\varphi_2$) (L>2).**

Now, we intend to demonstrate the phase coverage governed by the 1-bit time-varying phase modulation at different harmonic frequencies. **Figure 4a** depicts the results corresponding to different pairs of ($\varphi_1$, $\varphi_2$), while those presented in **Figure 4b** are restricted to $\varphi_1$=0° and $\varphi_2$=180°. When the acoustic elements are designed to expose $\varphi_1$=0° and $\varphi_2$=180° transmission phases, the equivalent phase coverage can reach $2\pi$ at any harmonic frequencies, except the central frequency, provided that the time-coding sequences are sufficiently long. The reason is attributed to the fact that the equivalent transmission coefficient at the zeroth harmonic is nothing but a simple average of exp(j0) and exp(jπ) factors. With the proper choice of ($\varphi_1$, $\varphi_2$), however, a full phase coverage can also be obtained at the central frequency. Note that the modulation period has no influence on the harmonic intensity but plays a vital role in determining the positions of the spectral lines. Besides, the transmission efficiency may not be strong for all possible choices. All in all, as a key aspect of the proposed time-varying coding approach, although the electromechanical actuation enables only a pair of transmission phases ($\varphi_1$, $\varphi_2$), the equivalent excitation obeying Eq. (4), can attain almost $2\pi$ phase coverage by suitably designing the time-coding sequences. In this way, a simple dual-state metasurface periodically time switched can be used to construct 1-bit or arbitrary multi-bit digital coding elements at high-order harmonic frequencies, opening up a plethora of interesting applications such as Doppler cloak and velocity illusion of a system moving in space[62]. While, in order to design non-modulated acoustic elements with higher-bit coding, such as 2-bit and 3-bit schemes, the electromechanical tunning mechanism would be very complicated and organizing the control unit would also be a challenge. Based on Eq. (7), the time-coding sequences 01/… and 10/… with arbitrary ($\varphi_1$, $\varphi_2$) are effective in construction of 1-bit coding elements at ±3rd and ±1st order harmonic frequencies. Referring to **Figures 3a-e**, the equivalent amplitudes for ($\varphi_1$, $\varphi_2$) = (0, 45°), (0, 90°), (0, 135°), and (0, 180°) are equal to 0.24/0.08, 0.45/0.15, 0.58/0.19, and 0.63/0.21, respectively, in which the numbers before and after slash correspond to ±1st and ±3rd harmonic components, respectively. Moreover, the square wave time-domain signals 01/… and 10/… with ($\varphi_1$, $\varphi_2$) = (90°, 180°) and 01/… and 10/… with ($\varphi_1$, $\varphi_2$) = (0, 90°) yield the proper set of four 2-bit coding elements for ±3rd and ±1st order harmonics, where the equivalent transmission amplitudes are 0.15 and 0.45, respectively. The difference in the equivalent amplitudes will offer a new degree of freedom for controlling the magnitude of the scattering patterns in the next section. A major remaining challenge, however, is that the phase and amplitude corresponding to each harmonic frequency is not independent [Eqs. (4), (7)] and so, attaining an arbitrary phase and amplitude at a certain harmonic frequency may not necessarily admit a solution. We will suggest an alternative way to resolve this issue in the next section.

## 3. Results and Discussion

Now, we construct the space-time-coding acoustic digital metasurface comprising of M×M supercells independently controlled by distinct time-modulated signals. Without loss of generality, we assume a 1D acoustic wavefront manipulation configuration in which each column of sixteen coding elements shares the same control voltage, which corresponds to the same phase status. To provide enough spatial resolution for acoustic wave manipulation, several identical elements are grouped into the half-wavelength supercell. We suppose the equivalent transmission coefficient of the $(u, v)^{th}$ supercell is $\varphi_{uv}^m(m, n)$ at the $m^{th}$ harmonic frequency. Under the normal incidence of acoustic plane waves, owing to a direct connection between different coding patterns and their radiation patterns, the far-field pattern of the metasurface can be expressed by

$$P_{far}^m(\theta,\phi) = f_e(\theta,\phi) \cdot \sum_{u=1}^{M}\sum_{v=1}^{M} a_{uv}^m \exp\left\{-i(k_c + mk_m)D\sin\theta\left[\left(u-\frac{1}{2}\right)\cos\phi + \left(v-\frac{1}{2}\right)\sin\phi\right]\right\} \quad (8)$$

where, $k_c=2\pi f_c/c_{air}$ and $k_m=2\pi f_m/c_{air}$, $D$ indicates the size of each supercell, $\theta$ and $\phi$ are the elevation and azimuth angles of an arbitrary direction, respectively, and finally, $f_e(\theta, \phi)$ is the pattern function of a supercell. As a well-known assumption, the element factor is considered as $f_e(\theta,\varphi)=\cos(\theta)$ during the calculation of the far-field patterns. We can calculate the scattering pattern of the coding metasurface at any harmonic frequencies via Eq. (8), where, by controlling the time-coding sequences of the individual elements, a set of complex transmission coefficients $a_{uv}^m$ is synthesized to design single- or multi-bit coding elements at different harmonic frequencies and shape the overall scattering signatures in both space and frequency domains. More specifically, via Eq. (4), we can synthesize the equivalent amplitude and phase excitations of all elements at a specific harmonic frequency. To show the nonlinear scattering manipulation capability of the proposed acoustic metasurface with 16 time-modulated columns, we exploited the same 2-interval 1-bit and 2-bit coding elements of the previous section to realize diverse space-coding sequences at different harmonic frequencies. Considering the pulse width of τ=10 ms, the modulation periods disclose that the −3rd, −1st, +1st, and +3rd order harmonics operate at the frequencies of (2.85, 2.95, 3.05, 3.15) kHz. The corresponding 2D polar patterns have been simulated by COMSOL Multiphysics software, and the normalized

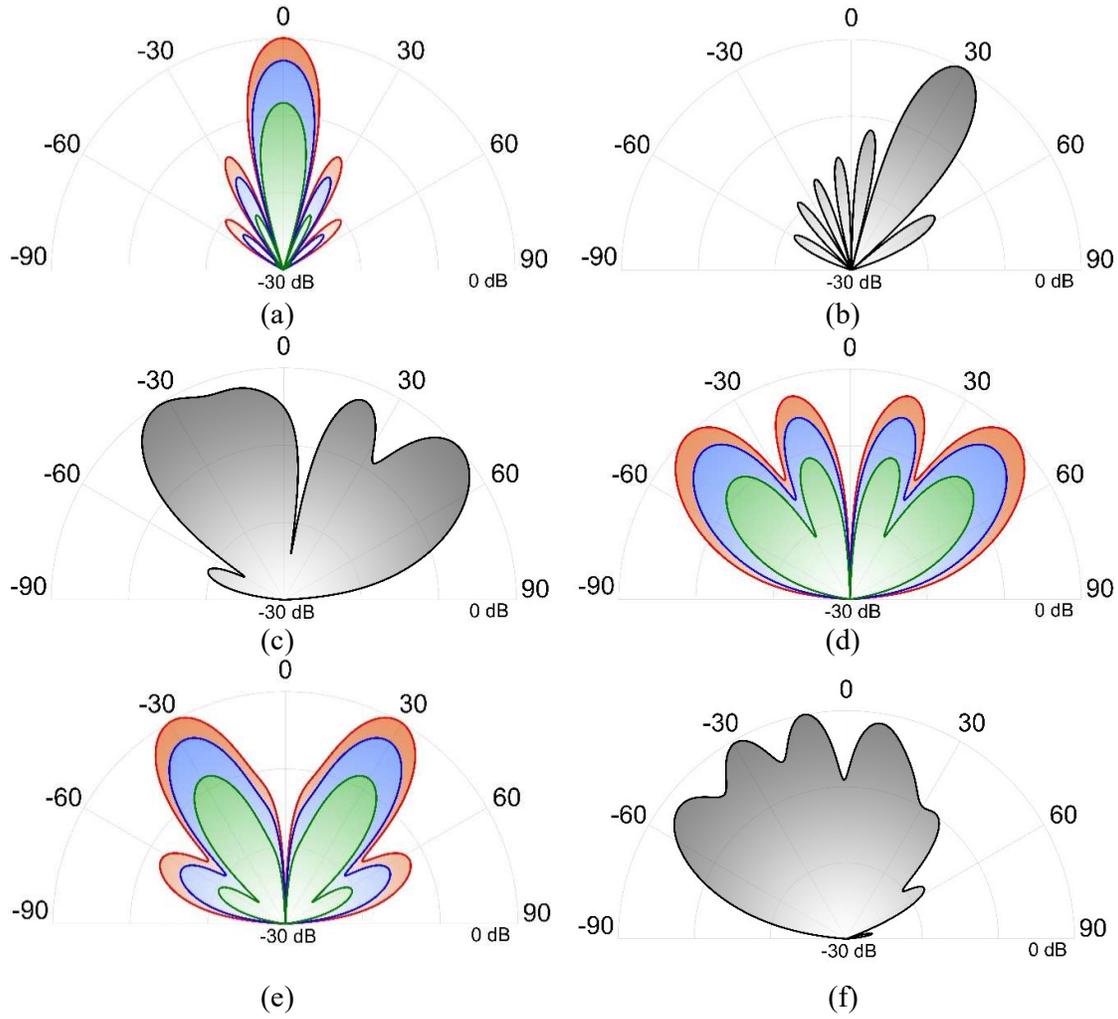

**Figure 5.** The calculated results of 2D polar patterns for different 1-bit and 2-bit space codes of (a) 00000000, (b) 01230123, (c) 13322012, (d) 10010011, (e) 11000110, and (f) 03331021 at different harmonics (a) -3rd, (b) +1st, (c) -1st, (d) +3rd, (e) -1st, and (f) +3rd order harmonic frequencies. The blue, red, and green linse in the transmission patterns of the 1-bit space codes stand for the time-domain control signals with ($\varphi_1$, $\varphi_2$) = (0, 45°), (0, 90°), and (0, 180°), respectively.

results in the far-field domain are demonstrated in **Figures 5a-f** for different sets of ($\varphi_1$, $\varphi_2$). As previously emphasized, the scattering magnitude of the far-field pattern for 1-bit space-coding sequences can be controlled somehow that by using ($\varphi_1$, $\varphi_2$) = (0, 45°) and (0, 90°) phase levels, the acoustic transmission patterns differ by −8.38 dB and −2.93 dB, at both ±1st and

±3rd harmonics, concerning those obtained by (0, 180°). In the case of space-coding sequence "00000000", a strong beam along the boresight direction is acquired for the −3rd order harmonic (**Figure 5a**). When the acoustic metasurface is programmed with space-coding sequences "10010011" (**Figure 5d**) and "11000110" (**Figure 5e**), due to the destructive acoustic interference between different columns, two and four main beams along specific angular directions emerge for the +3rd and −1st order harmonics, respectively. The 2-bit space modulation, however, provides more degrees of freedom whereby the acoustic metasurface driven by the space-coding sequences "01230123" (**Figure 5b**), "13322012" (**Figure 5c**), and "03331021" (**Figure 5f**) produce single-beam, triple-beam, and multiple-beam far-field patterns at +1st, −1st, and +3rd harmonic components, respectively. The codes '0', '1', '2', '3', and '4' are a brief representation of '00', '01', '10', '11', respectively. The sound far-field patterns corresponding to the 1-bit space codes driven by the time-domain signals ($\varphi_1$, $\varphi_2$) = (0, 45°) and (0, 90°) are also given in **Figures 5a**, **d**, and **e** in red and green from which, in accordance with the analytical interpretations, the scattering magnitudes can be elaborately adjusted without altering the spatial profile over the surface.

Until now, we have only considered the feasibility of acoustic harmonic generation through the time-domain digital coding metasurface with 2-interval control signals. The presented time modulation scheme has several shortcomings: 1) there is a strong correlation between the harmonic amplitude and phase, thereby greatly hindering independent control of both parameters in practice; 2) the acoustic wave manipulation is enabled only for odd-order harmonics; and finally, 3) designing multi-phase digital elements with large carrier-to-harmonic conversion rate is still a challenge. More flexibility can be obtained by increasing the length of the time-coding sequence. Exploiting the time-coding sequences with L>2 unlocks accessing to the acoustic wave manipulation at the even harmonics and provides higher equivalent transmission amplitudes. A new solution for prevailing over the inherent phase/amplitude coupling is to impose an additional time delay $t_0$, to the time-coding sequence. According to the time shift property of Fourier transform with $F[\xi_{uv}(t-t_0)]=\xi_{uv}(\omega)\exp(-j\omega t_0)$, the time delay $t_0$ creates an additional phase shift $-2m\pi t_0/T_m$ for the $m^{th}$ order harmonic while the corresponding amplitude remains unchanged. It is worth mentioning that the time delay does not affect the transmission properties at the central frequency. The recipe for obtaining multi-bit coding elements at a pre-determined frequency harmonic is thus to seek for only one proper time-coding sequence, which yields high equivalent transmission amplitude at the same spectral position. Subsequently, the required phase shifts for designing 1-bit (phase step $\pi$), 2-bit (phase step $\pi/2$), and 3-bit (phase step $\pi/3$) digital elements at the $m^{th}$ order harmonic are simply acquired via introducing the time delays $t_0=T_m/2m$, $t_0=T_m/4m$, and $t_0=T_m/8m$ to the time-coding sequences, respectively. The optimum time-coding sequences are obtained as "11010101", "10010110", "10011011", and "00111100" for ±4th, ±3rd, ±2nd, and ±1st order harmonic frequencies, respectively. With a pulse width of 10 ms, the modulation frequency of the system is 12.5 Hz. **Figures 6a-d** display the contribution of different harmonic frequencies to the overall energy passing through the acoustic metasurface. Following this figure, 44%, 50%, 50%, and 81% of the total transmitted energy is successfully dedicated to ±4th, ±3rd, ±2nd, and ±1st order harmonic frequencies, respectively. To inspect the nonlinear performance of the proposed space-time-coding acoustic metasurface, the scattering signatures of different 1-bit, 2-bit, and 3-bit space-coding sequences are investigated. The far-field patterns pertaining to the 1-bit, 2-bit, and 3-bit space-coding sequences are shown in **Figures 7a-f**, **Figures 8a-f**, and **Figures 9a-f**, respectively. The angular direction of the acoustic beams emitted via the gradient coding sequences can be theoretically predicted based on the generalized Snell's law[52,63,64]

$$\theta_r^m = \sin^{-1}\left\{\frac{c_{air}}{2\pi(f_c+mf_m)}\frac{d\phi}{dx}\right\} \quad (9)$$

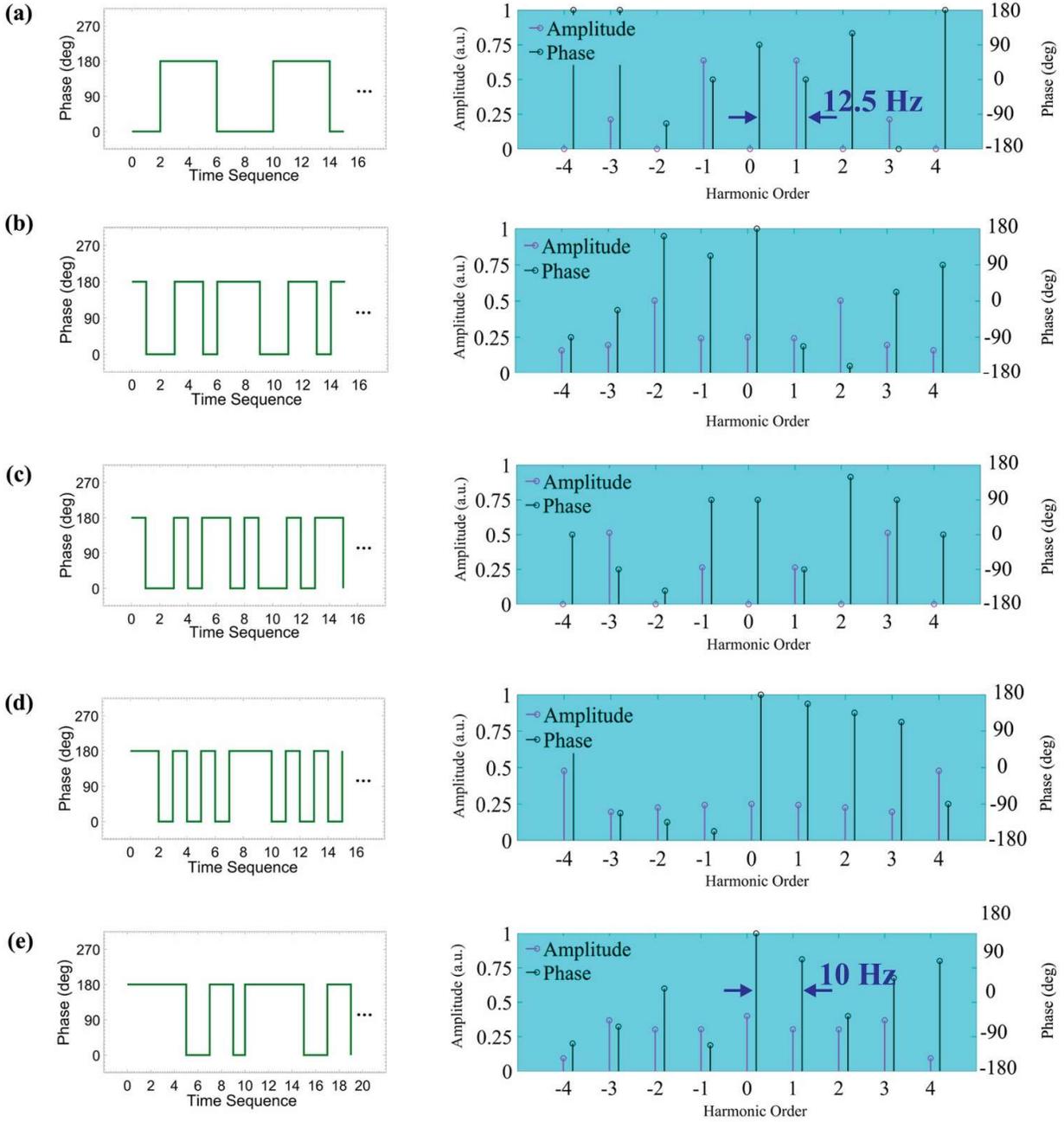

Figure 6. The harmonic phases/intensities distribution of the proposed time-domain digital acoustic metasurface at 3 kHz with the pulse duration $\tau$=10 ms when the transmission phase of the acoustic elements is periodically switched between 0 and $\pi$ under the optimized time-varying coding sequences of (a) 00111100, (b) 10011011, (c) 10010110, (d) 11010101, and (e) 1111100110 for maximum possible coupling of the carrier power to (a) ±1st, (b) ±2nd, (c) ±3rd, (d) ±4th order harmonics or (e) distributing it equally between these components.

It should be noted that due to the fact that the time-delay phase shifts corresponding to the $+m$ and $-m$ harmonics differ only in a sign, the far-field pattern at $f_c+mf_c$ is the mirror of that appears at $f_c-mf_c$ with respect to $\theta=0$. Therefore, the scattering patterns are plotted for one of them. Upon programming by the space codes "00000000" (**Figure 7a**), "00110011" (**Figure 7b**), and "00001111" (**Figure 7c**), the acoustic metasurface refracts the normally incident wave into $\theta=0°$ (2nd order harmonic), $\theta=\pm 29.1°$ (1st order harmonic), and $\theta=\pm 13.8°$ (−1st order harmonic) directions, respectively. The numerical results have an excellent agreement with the calculated angles, *i.e.*, 0, 30°, and 14.47°. The slight discrepancies are due to the angular dependency of the element pattern function. As an unbalanced gradient sequence, the space code "00100100" has a DC component due to the different numbers of 0 and 1 digits. Therefore, following the Fourier connection between the space-coding sequence and the acoustic far-field pattern, an additional pencil beam pointing at the boresight direction is also attained, yielding a three-beam scattering pattern (**Figure 7d**). Finally, a random space code "11010010" is utilized to demonstrate the ability of the 1-bit coding layouts in re-distributing four beams, symmetrically, on the other side of the metasurface (**Figure 7e**). As proved, the time modulation also enables 2-bit and 3-bit space modulation at any desired harmonic frequency. Let us assume that the sequence $G_n$ indicates a gradient space code "01230213," in which $n$ is an integer indicating the repetition number. For instance, $G_2$ is "01230213". Following the generalized Snell's law in Eq. (9) and the half-wavelength size of the acoustic supercells, a 2-bit gradient acoustic metasurface, can generate a pencil beam with a few numbers of available tilt-angles $\sin^{-1}(1/2n)$, when the integer number equals n=1, 2, 3, …. For example, the time-modulated acoustic metasurface endowed with $G_1$, $G_2$, and $G_3$ space codes generate a pencil beam oriented along $\theta=30°$, $\theta=14.47°$, and $\theta=9.59°$ directions. The far-field patterns related to the $G_1$ and $G_2$ sequences are shown in **Figure 8c** (at –1st harmonic) and **Figure 8a** (at –4th harmonic), respectively, where the numerical tilt-angles depict a good agreement with the calculated ones. To enhance the manipulating abilities of the coding metasurface, significantly, inspired by the frequency-shift property of the Fourier transform, we utilize a principle called the scattering-pattern shift[65] using the convolutional theorem through the modulus of two different space-coding sequences. By the modulus of different sets of gradient space codes, for instance, $G_1 \circledast G_2$ and $G_1 \circledast G_{-4}$, the mixed sequences, *i.e.,* "01302312" (**Figure 8d,** +3rd order harmonic), and "3012230112300123" (**Figure 8b,** +2nd order harmonic), yield a single-beam transmission pattern pointing at $\theta=47.9°$ and $\theta=21.7°$, respectively. The numerical tilt-angles corroborate well the theoretical values calculated by $\theta=\sin^{-1}(\sin\theta_1+\sin\theta_2)$, in which $\theta_1$ and $\theta_2$ stand for the beam orientation of two sequences being mixed by using the convolution operation[65]. Note that as achieving a larger refraction angle requires space code with a faster variation rate, emerging parasitic scattering beams is inevitable (see **Figure 8d**) since the 2-bit modulation is still associated with an abrupt phase change for the neighbouring cells. The convolution principle can be further applied to reach flexible and continuous control of acoustic wavefronts. An asymmetric multi-beam far-field pattern can be elaborately obtained by 2-bit space codes. **Figure 8e** better elucidates the scattering-pattern shift property of the coding metasurfaces in which the corresponding space code "01230123123012300" is achieved by the modulus of a 2-bit [$G_2$ (single beam)] sequence and a 1-bit ["0000000011111111" (dual beam)] code. The yielded transmission pattern, which is demonstrated for the +4th order harmonic, has two pencil beams locating at the left-side half-space, just around the specific direction associated with $G_2$ code. The access to multi-bit acoustic elements, which is provided by the proposed time modulation scheme, allows us to apply more advanced operations on the digital space codes at different harmonic components. The universal theory for addition operation implies that any superposition connection between two N-bit complex codes results in an (N+1)-bit complex code whose far-field pattern includes the main beams of both contributing space codes at the same time[66,67]. Beginning from 1-bit codes, we performed addition operations of two dual-beam space codes 00001111⊕00110011 =00113322, whereby the superimposed metasurface refracts the incident acoustic wave along four main beams with $\theta_{1,2}$=13.8° and $\theta_{2,3}$=±29.4° directions. **Figure 8f** displays the corresponding transmission pattern for −3rd harmonic frequency, and the superimposed space-coding sequence is calculated based on the general regulations for addition operations[66]. Since the superposition of two distinct 2-bit sequences will generally yield a 3-bit code, by using 3-bit space-coding sequences, more complicated scattering signatures can be achieved. Besides, with identical tilt-angles, the anomalous transmission resulted from the 3-bit gradient space codes is accompanied by lower parasitic beams due to the smoother phase variations over the surface. **Figures 9a-f** illustrate several advanced scattering functionalities achievable by the 3-bit space-coding sequences "01234567", "72107210", "0167123023457456", "761103325547", "74451267", and "63515172", each of which is demonstrated at a specific harmonic. The space codes in **Figures 9b-e** are obtained by the superposition principle and convolutional theorem. As another interesting application, especially in single-source acoustic imaging, when a maximally-randomized space-coding sequence is employed, the monochromatic sound is transmitted and dispersed in multiple spatial directions after being converted to a specific harmonic frequency. This is confirmed by **Figure 9f**, in which the random coding sequence "63515172" is determined by accomplishing the same optimization procedures originally introduced for the electromagnetic diffusers[68,69].

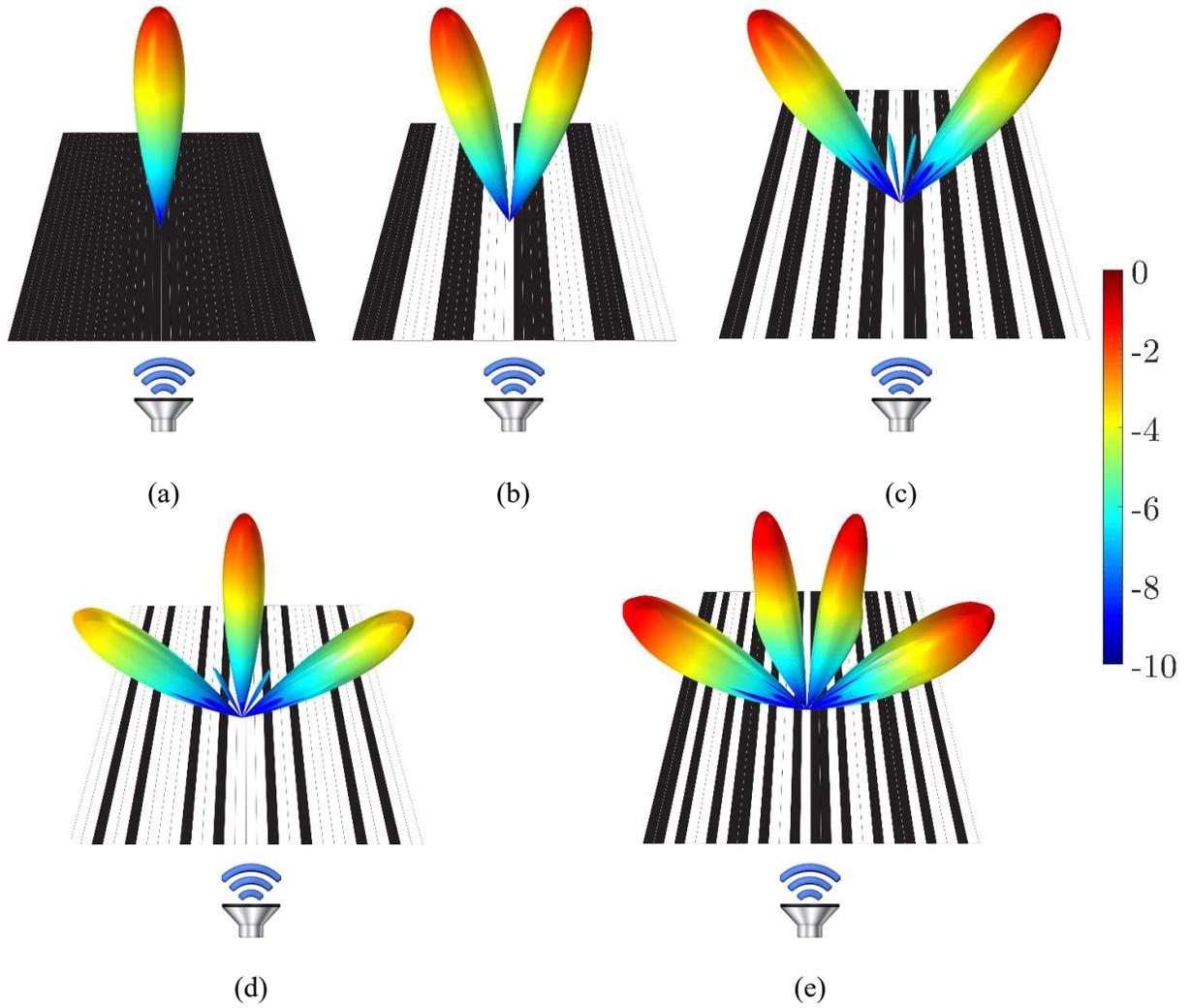

**Figure 7.** Numerical performance of the proposed space-time-coding acoustic metasurface with different 1-bit space codes of (a) 00000000, (b) 00001111, (c) 00110011, (d) 00100100, and (e) 11010010 at different harmonics (a) +2nd, (b) +1st, (c) -1st, (d) -3rd, and (e) -2nd order harmonic frequencies.

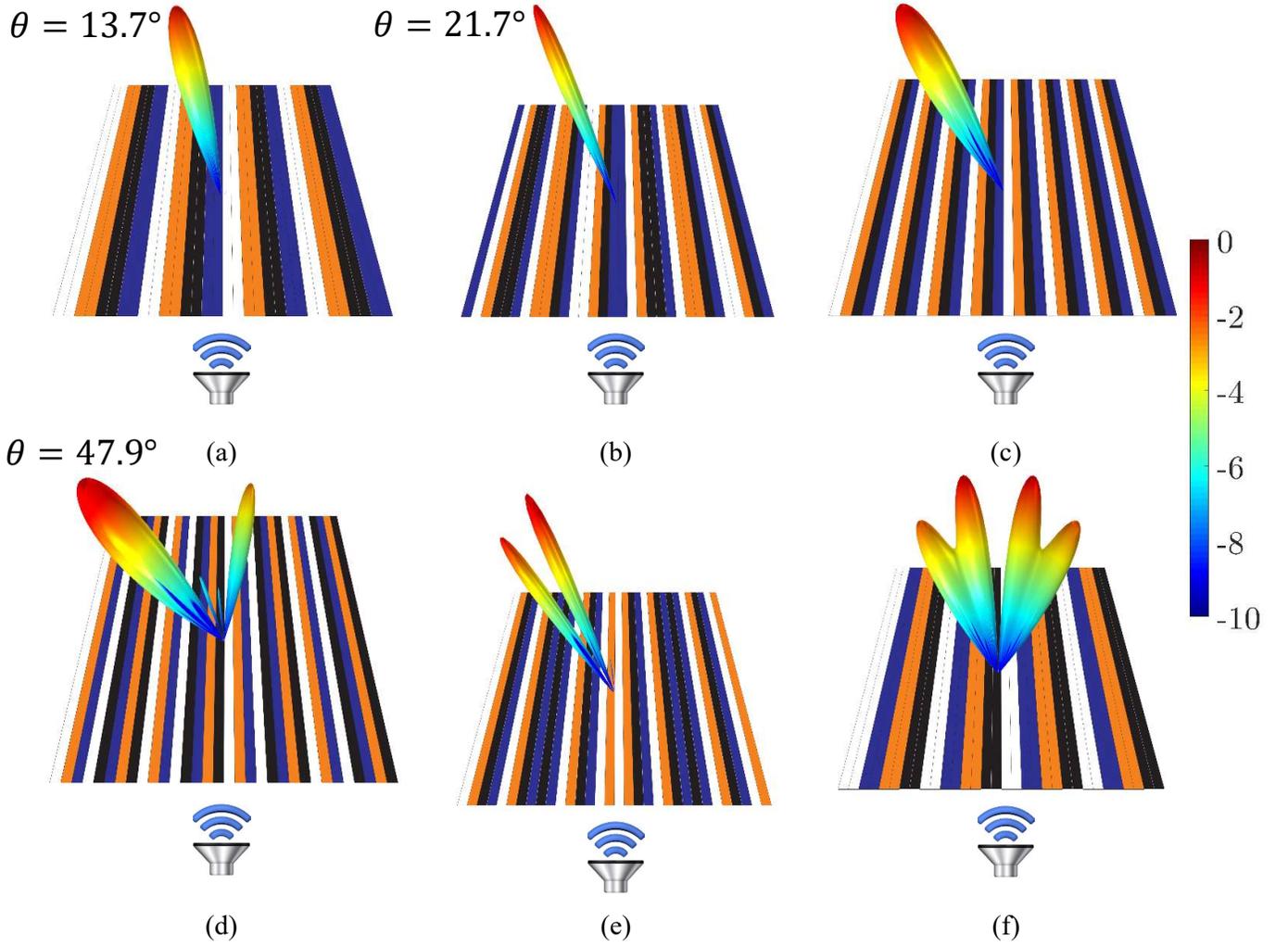

**Figure 8. Numerical performance of the proposed space-time-coding acoustic metasurface with different 2-bit space codes of (a) 00112233 ($G_2$), (b) 3012230112300123 ($G_1 \odot G_{-4}$), (c) 01230123 ($G_1$), (d) 01302312 ($G_1 \odot G_2$), (e) 0123012312301230 ($G_1 \circledast 0000000011111111$), and (f) 00113322 (00001111$\oplus$00110011) at different harmonics (a) -4th, (b) +2nd, (c) -1st, (d) +3rd, (e) +4th, and (f) -3rd order harmonic frequencies.**

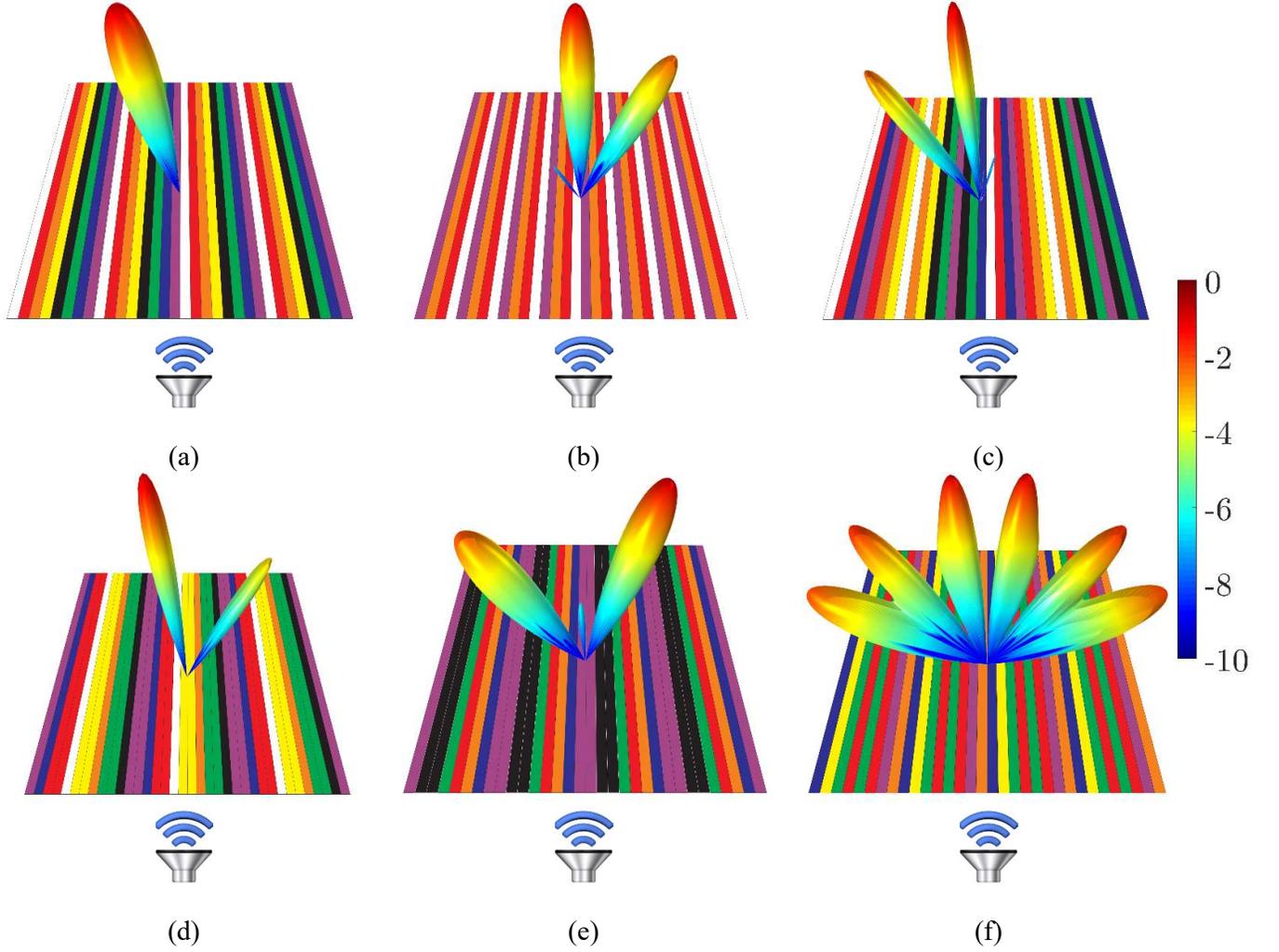

**Figure 9. Numerical performance of the proposed space-time-coding acoustic metasurface with different 3-bit space codes of (a) 01234567, (b) 72107210, (c) 0167123023457456, (d) 761103325547, (e) 74451267, and (f) 63515172 (random) at different harmonics (a) -2nd, (b) +4th, (c) -1st, (d) +1st, (e) +2nd, and (f) +3rd order harmonic frequencies.**

As an important deduction from this wave manipulation tour, the space-time-coding digital metasurface is capable of generating diverse acoustic scattering patterns at the desired harmonic frequency, thus removing the need for intricate phase-shift networks[70]. In all the above demonstrations, the space-coding sequences can be realized in different harmonic components, provided that the suitable time-coding sequence is applied. The above-mentioned method hints toward a novel route for controlling the scattering patterns of one specific harmonic frequency by breaking the constraint between their amplitude and phase. This means that the other harmonics have uncontrolled orientations under the same coding sequences. Now, we intend to demonstrate the ability of the proposed space-time-coding strategy to manipulate the beam direction pertaining to multiple harmonic frequencies, which is hard to achieve using the conventional methods. For better equalization of the power levels, we exploit a binary optimization to find the best time-coding sequence of length L=10, which spreads the power equally between the desired harmonics (**Figure 6e**). If the time-coding sequences of the adjacent columns have a time delay equal to $t_0$, an additional space phase shift $\Delta\varphi^m = 2\pi m t_0 / T_m$ between $(u, v)^{th}$ and $(u, v+1)^{th}$ elements appears. Upon illuminating by a normal incidence, the generalized Snell's law can be expressed as

$$\theta_b^m \simeq \sin^{-1}\left(\frac{\lambda_c}{2\pi}\frac{d\phi^m}{dx}\right) = \sin^{-1}\left(\frac{\lambda_c}{2\pi}\frac{\Delta\phi^m}{D}\right) = \sin^{-1}\left(m \times \frac{\lambda_c}{D} \times \frac{t_0}{T_m}\right) = \sin^{-1}\left(2m \times \frac{t_0}{T_m}\right) \quad (10)$$

The 2D and 3D scattering patterns along with the time-domain sequences, are displayed in **Figures 10a, b**, by introducing the time delay $t_0 = 1/8\ T_m$. The fundamental coding element is obtained via the time-domain controlling signal "1111100110" for which the harmonic frequencies m=–3 to m=+3 have identical transmission amplitudes in the range of 0.3<|a|<0.4. The modulation period $T_m$ of the time-coding sequences is 100 ms that corresponds to a system modulation frequency $f_m$ =10 Hz. It can be observed that the main beams at different harmonic frequencies point to different directions ($\theta = \pm47.9°, \pm29°, \pm14.1°,$ and $0°$), thereby realizing the desired harmonic beam steering. As can be seen from 2D far-field patterns, the calculated ($\sin^{-1}[0.25m]$) and simulated beam orientations are in an excellent agreement. We should remark that although the examples in this paper are presented for 1D scenarios, the proposed time-varying coding strategy can be simply extended to the 2D acoustic metasurfaces. In this case, each contributing element is independently controlled with its own square wave signal so that the other intriguing applications such as acoustic diffusion, acoustic OAM generation, and acoustic holograms can also be acquired.

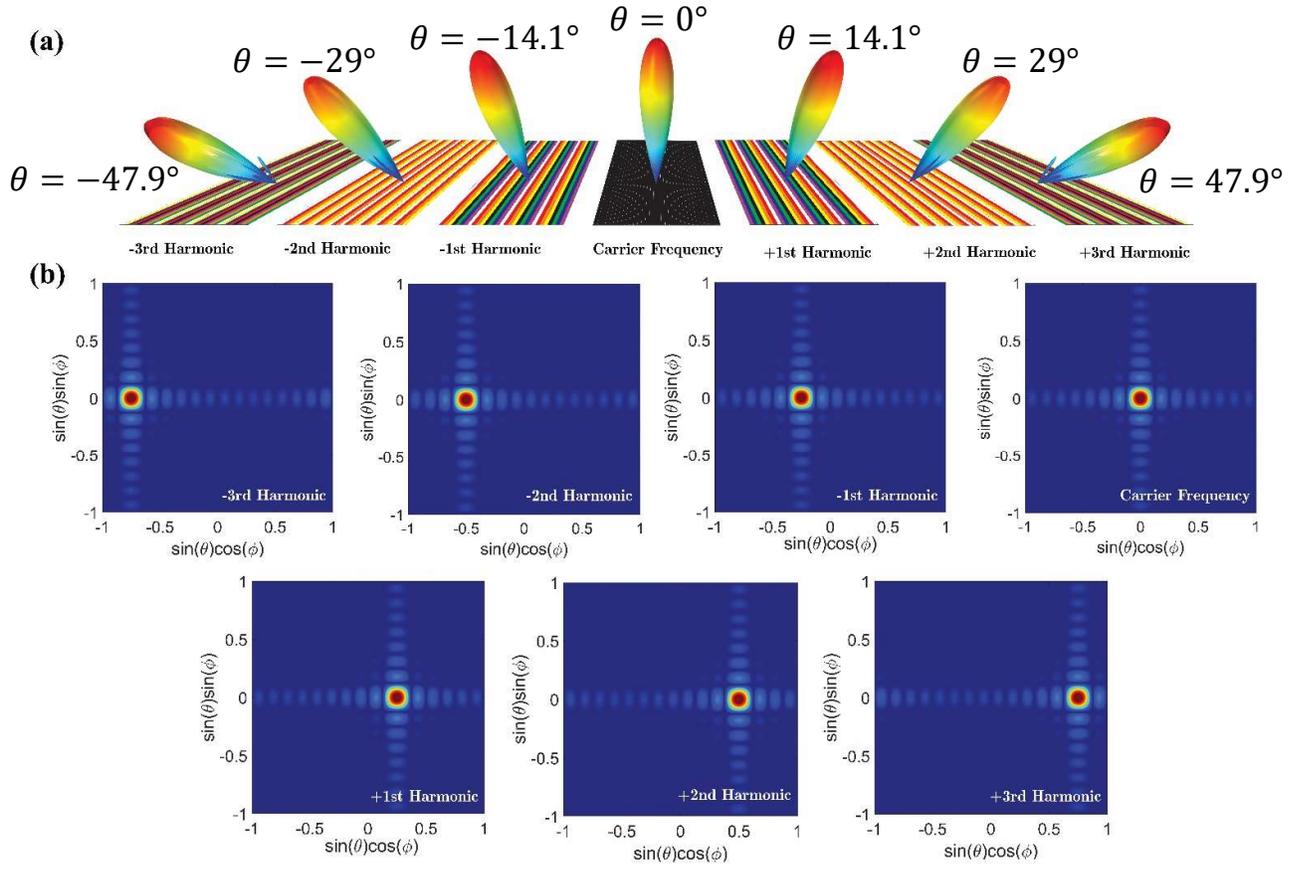

**Figure 10.** Demonstrating the capability of the proposed space-time-coding acoustic metasurface in control of multiple harmonic frequencies at the same time by introducing suitable space-dependent time-delays to the time-coding sequences. The space codes are 03614725, 01230123, 01234567, 00000000, 76543210, 32103210, and 52741630 (from left to right). (a) 3D and (2) 2D far-field patterns are plotted for different harmonic orders scanning the whole upper space. The monochromatic sound impinges on the metasurface from the bottom side.

## 4. Conclusion

In conclusion, the simultaneous manipulation of the acoustic waves in both spatial (far-field pattern) and frequency (frequency distribution) domains has been demonstrated by introducing the space-time-coding digital acoustic metasurfaces for the first time. The programmable acoustic elements are based on the Helmholtz resonators whose transmission phase responses can be dynamically tuned via an external electromechanical system fed by time-coding sequences switched cyclically in a modulation period. With independent and accurate control of the harmonic amplitudes/phases, several illustrative examples have been presented to demonstrate the unique beamforming capability of the designed space-time-modulated acoustic metasurface at different harmonic frequencies. The numerical simulations have excellent conformity with our theoretical predictions. The present work enriches the functionalities of the conventional acoustic metasurfaces, drastically, and may find intriguing applications, such as acoustical illusion apparatuses.